\documentclass[preprint,aps,superscriptaddress,floatfix]{revtex4}
\usepackage{graphicx,epsfig,amssymb,amsmath,array}
\usepackage{relsize,placeins,float}
\usepackage[dvipsnames]{xcolor}
\usepackage{blindtext}
\usepackage[english]{babel}

\begin{document}

\title{Transmitted resonance in a coupled system}

\author{Mattia Coccolo}
\affiliation{Nonlinear Dynamics, Chaos and Complex Systems Group, Departamento de F\'{i}sica,
Universidad Rey Juan Carlos, Tulip\'{a}n s/n, 28933 M\'{o}stoles, Madrid, Spain}

\author{Miguel A.F. Sanju\'{a}n}
\affiliation{Nonlinear Dynamics, Chaos and Complex Systems Group, Departamento de F\'{i}sica,
Universidad Rey Juan Carlos, Tulip\'{a}n s/n, 28933 M\'{o}stoles, Madrid, Spain}

\date{\today}
\date{\today}

\begin{abstract}

When two systems are coupled, one can play the role of the driver, and the other can be the driven or response system.  In this scenario, the driver system can behave as an external forcing. Thus, we study its interaction when a periodic forcing drives the driver system. In the analysis a new phenomenon shows up:  when the driver system is forced by a periodic forcing, it can suffer a resonance and this resonance can be transmitted through the coupling mechanism to the driven system. Moreover, in some cases the enhanced oscillations amplitude can also interplay with a previous resonance already acting in the driven system dynamics. 
\end{abstract}

\maketitle

\section{Introduction}

Driven systems have had relevance in quite a few scientific fields such as medicine \cite{jiruska,mormann,Rulkov}, physics \cite{jensen,hramov}, communication \cite{Koronovskii,Naderi}, mechanics \cite{defoort}, network \cite{delellis,zhang}, circuits \cite{Yao} and engineering \cite{sujith}, among others. In fact, their behaviors can explain the phase transitions in ferromagnetic materials, the dynamics of which were studied using mean field coupling \cite{Desai,Dawson}. Moreover, Winfree \cite{Winfree} proposed to model the phenomenon of collective synchronization in terms of populations of coupled nonlinear oscillators, where each one has its own globally stable limit cycle. The link between synchronous flashing of the fireflies and the origin of the brain rhythms was established, on a qualitative level, in \cite{Wiener} in a neuroscience realm. Finally, an interesting aspect is to study when these dynamics are not just synchronized but enhanced by the coupling mechanism, due to the emergence of a resonance.

The importance to study resonance phenomena can be found in the widespread literature. As a matter of fact, different kinds of nonlinear resonances have been broadly studied such as stochastic resonance \cite{Gammaitoni, McDonnell}, chaotic resonance \cite{Zambrano},  vibrational resonance \cite{Landa}, delay-induced resonance \cite{Cantisan,Coccolo} and Bogdnaov-Takens resonance \cite{Coccolo2}. Each of these take their names by the mechanism that induce them. Furthermore, the resonance phenomenon holds a crucial significance in coupled systems, as it actively boosts the overall performance and efficiency of the system. 

These ideas intersect with various scientific fields, including acoustics, where the interconnected strings of a musical instrument parallel the coupling seen among planets in astrophysics. Furthermore, diverse engineering disciplines delve into coupling within different contexts, exploring interactions among various components in engines or buildings.

In a recent study (\cite{Coccolo_synch}), the authors investigated the impact of a time-delayed driver system serving as the sole external forcing on the oscillation amplitude of a non-delayed driven system. Their findings revealed the emergence of a resonance triggered by the coupling mechanism. The results of that article constitute the initial point from which we start our analysis. We proceed to examine the impact of the interaction between the time-delayed driver system, acting as an external forcing, and a periodic forcing on the dynamics of the driven system, called the {\it response system}. This interaction can lead to abrupt changes in the oscillator behavior. Therefore, we use the coupling mechanism of the {\it continuous control} \cite{Ding,Kapitaniak}. This method couples the two systems through the coupling constant, which is a parameter of the response system that multiplies the difference between the signal of the driven system and the response system. Other methods might be used to couple two systems \cite{Pecora1,Pecora2,Pecora3}, nevertheless we consider this models in the easiest way the driver system effects on the response system as an external forcing.

After establishing the coupling method, we apply it to examine a specific phenomenon. This phenomenon involves the transmission of the resonance in the driver system, initiated by its external periodic forcing, to the response system through the established coupling mechanism.

As the reader can realize, to accomplish our goal, we had to advance carefully in our research due to the extensive number of parameters that can be varied. So our work plan has been to vary the amplitude of the external periodic forcing.  This has been done by fixing the coupling constant value $C=1.66$, that has been proved to be significant in our previous article \cite{Coccolo_synch}. Then, we analyze the effect of varying the coupling constant in function of the amplitude of the response system periodic forcing. 

To summarize, we employ a periodic forcing to induce resonance within the driver system and subsequently transmitting it to the response system via the coupling mechanism. In other words, we analyze the interaction between the coupling mechanism and the periodic forcing that drives the driver system. 

The structure of this paper is as follows. After discussing the model in Sec.~\ref{Sec:II}, we analyze the specific phenomenon under investigation by showing the most interesting results. In Sec.~\ref{Sec:III}, we analyze the transmission of the driver system resonance to the response system at a fixed coupling constant value, while in Sec.~\ref{Sec:IV}, we examine the impact of varying the coupling constant value on the resonance phenomenon. Finally, in Sec.~\ref{Sec:concl} some concluding remarks are presented.

\section{Model}\label{Sec:II}
We study the response of a Duffing oscillator (the response system) when it is driven by a time-delayed Duffing oscillator (the driver system).  The two systems are coupled with a simple form of unidirectional coupling, the {\it continuous control}  \cite{Ding,Kapitaniak}. This coupling mechanism can be summarized in the following way.  

\begin{figure}[htbp]
  \centering
   \includegraphics[width=12.0cm,clip=true]{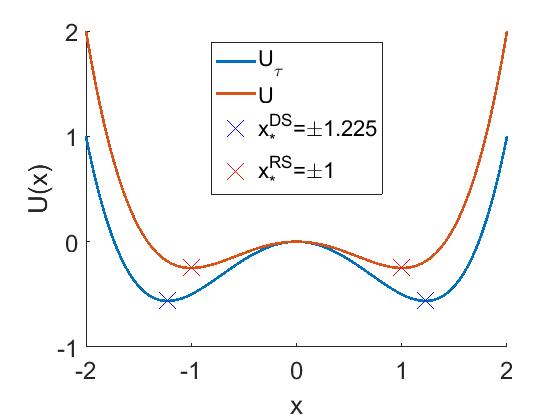}
   \caption{The double-well potential of the Duffing oscillator and of the time-delayed Duffing oscillator indicating the stable fixed points. The blue curve and the blue crosses are referred to the time-delayed Duffing oscillator, the red curve and the red crosses to the Duffing without delay.}
\label{fig:1}
\end{figure}

\begin{figure}[htbp]
  \centering
   \includegraphics[width=15.0cm,clip=true]{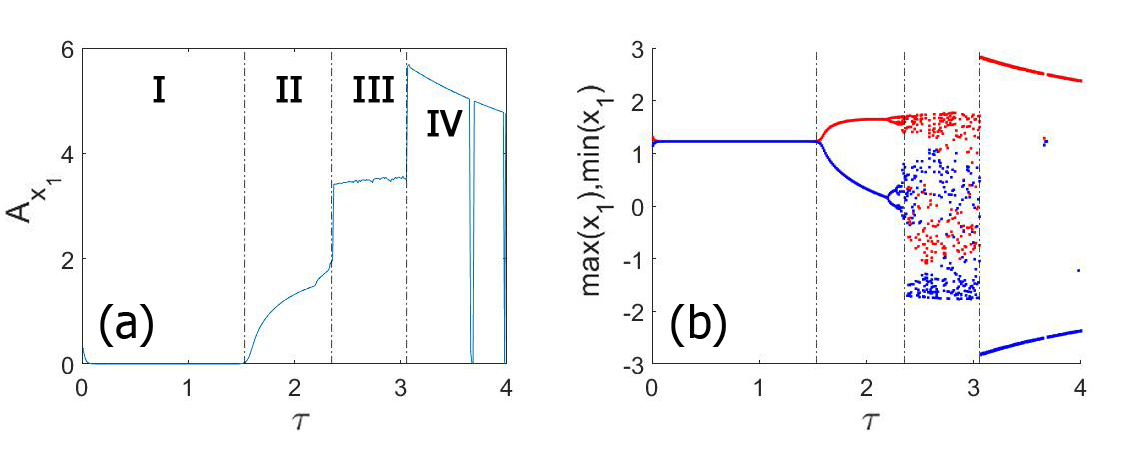}
   \caption{The figure shows the oscillatory behavior of the driver system for all $\tau$ regions. In panel (a) we show the oscillations amplitude $A_{x_1}$ and in panel (b) the maximum-minimum diagram. The history functions for the driver system are constant $(u_{0},v_{0})=(1,1)$.}
\label{fig:2}
\end{figure}

Given a square matrix $C$ of dimension $n$ with constant elements, the matrix $C$ multiplies the vector of differences $(\mathbf{x_1} - \mathbf{x_2})$, where $\mathbf{x_1} \in \mathbb{R}^n$ represents the vector of variables for the driver system, and $\mathbf{x_2} \in \mathbb{R}^n$ represents the vector of variables for the response system. In our case, the matrix $\mathbf{C}$ has only one nonzero element that is the coupling constant $C$ that multiplies the difference of the variables $(x_1-x_2)$, related with the two oscillators as explained before.  The numerical value of the coupling constant  measures the strength of the coupling for the forcing induced by the time-delayed Duffing oscillator. Moreover, one external periodic forcing show up in the equations of the systems, yielding 
\begin{align}\label{eq:1}
Driver&\rightarrow \quad   \frac{d^2x_1}{dt^2}+\mu\frac{dx_1}{dt}+\gamma x_1(t-\tau)+\alpha x_1(1-x_1^2)=F\cos{\omega_1 t},\\
Response&\rightarrow \quad   \frac{d^2x_2}{dt^2}+\mu\frac{dx_2}{dt}+\alpha x_2(1-x_2^2)=C(x_1-x_2), 
\end{align}
where we have fixed the parameters $\mu=0.01$, $\alpha=-1$. Certainly, the feedback gain $\gamma$ of the time-delayed term plays a crucial role in shaping the system dynamics, as highlighted in \cite{Jin1,Jin2,Jin3}.  Exploring the effects of different feedback gains undoubtedly could offer valuable insights. However, we have considered fixing $\gamma=-0.5$ since the general validity of our conclusions is not affected. Furthermore, this choice ensures the clarity in the analysis of the resonance phenomenon. The  damping parameter $\mu$ is kept small in order to better appreciate the effects of the variation of the external forcing, the driver system and the periodic one, on the dynamics of the response system and the correlation with the coupling constant $C$. The history functions of the driver system are $u_0=v_0=1$, and the initial conditions of the response system are $x_0=y_0=0.5$.
The potentials and the fixed points of both systems are shown in Fig.~\ref{fig:1}, where the unstable fixed point $x_0=0$ is the same for the two potentials. Conversely, the stable fixed points of the driver system are
\begin{equation}\label{eq:2}
x^{DS}_{*}=\pm\sqrt{\frac{\alpha+\gamma}{\alpha}}=\pm1.225,
\end{equation}
and the stable fixed points of the response system are
\begin{equation}\label{eq:3}
x^{RS}_{*}=\pm\sqrt{\frac{\alpha}{\alpha}}=\pm1.
\end{equation}
As it has already been reported in \cite{Coccolo, Cantisan}, the time-delayed Duffing oscillator  without forcing undergoes various bifurcations while $\tau$ changes. We show its behavior in Fig.~\ref{fig:2}(a-b). In the first one, we plot the oscillations amplitude and in the second one the maximum-minimum diagram of the oscillations. This last one has been plotted by representing on the figure the maxima and minima of the last 5 periods of the oscillations for each value of $\tau$.  Four regions are discernible in the figures although only the first two regions are of interest for our purposes. In fact, the third one $2.35<\tau<3.05$ is chaotic and our study shows that its analysis does not contribute to the goal of this paper. In the fourth one, $\tau>3.05$, trajectories are periodic and are already no longer confined to either one of the wells. Moreover, the introduction of an external periodic forcing does not alter the response system oscillations induced through the coupling mechanism. Then, in the first region for $\tau<1.53$ the oscillations fall on one of the fixed point. Interestingly  for $\tau\lesssim0.1$, the time for the oscillations to go to the fixed point is larger for smaller values of $\tau$.  When $\tau\rightarrow0$ that time goes asymptotically to $\infty$. Next, in the second region $1.53<\tau<2.35$ the oscillations are sustained but confined to one of the wells. Therefore, it is feasible to observe the emergence of a resonance in these two regions.\\ 
Throughout the article, we denote the variable of the driver system as $x_1$ and the variable of the response system as $x_2$.  Furthermore, in the following section, we set $C$ equal to $1.66$ value that leads to an increase in the amplitude of oscillations in the response system, even in the absence of an external periodic forcing. This assures us that any further enhancement is due to the new phenomenon and not as a by-product of the cooperation between the coupling constant and the dynamics of the system.

\section{Transmitted resonance}\label{Sec:III}

\begin{figure}[htbp]
  \centering
   \includegraphics[width=15.0cm,clip=true]{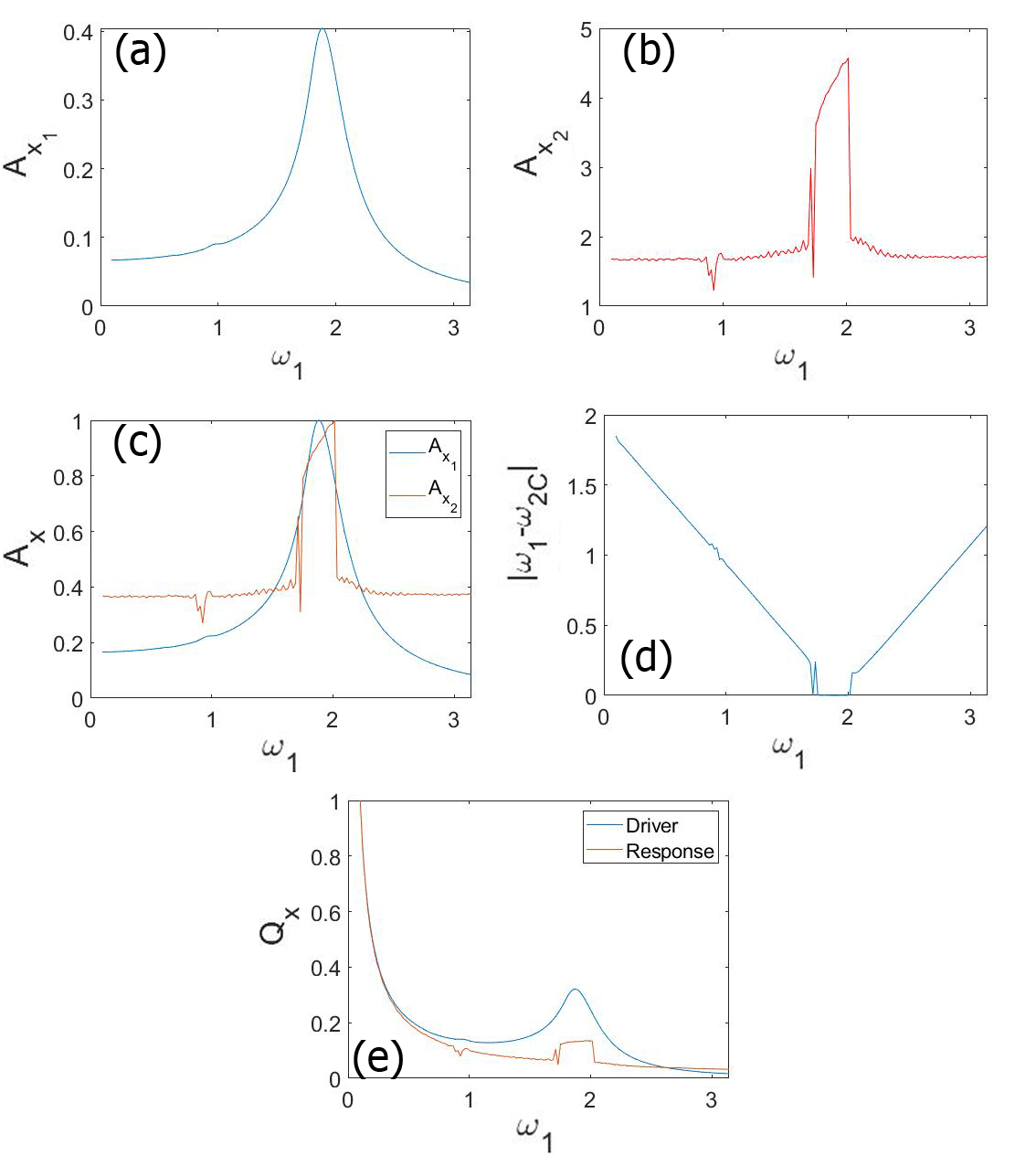}
   \caption{ The figure shows in panel (a) the oscillations amplitude of the driver system, and  in panel (b) the oscillations amplitude of the response system for $F=0.1$ and $\tau=1$. Panels (c) and (d) are a comparison of the oscillations amplitude and frequencies of the two oscillators, respectively. The frequency of the driver is the forcing frequency while the response system is calculated with the Fast Fourier Transform (FFT) and we have called it $\omega_{2C}$. The last panel (e) shows the $Q-$factor for the two systems. }
\label{fig:3}
\end{figure}

\begin{figure}[htbp]
  \centering
   \includegraphics[width=15.0cm,clip=true]{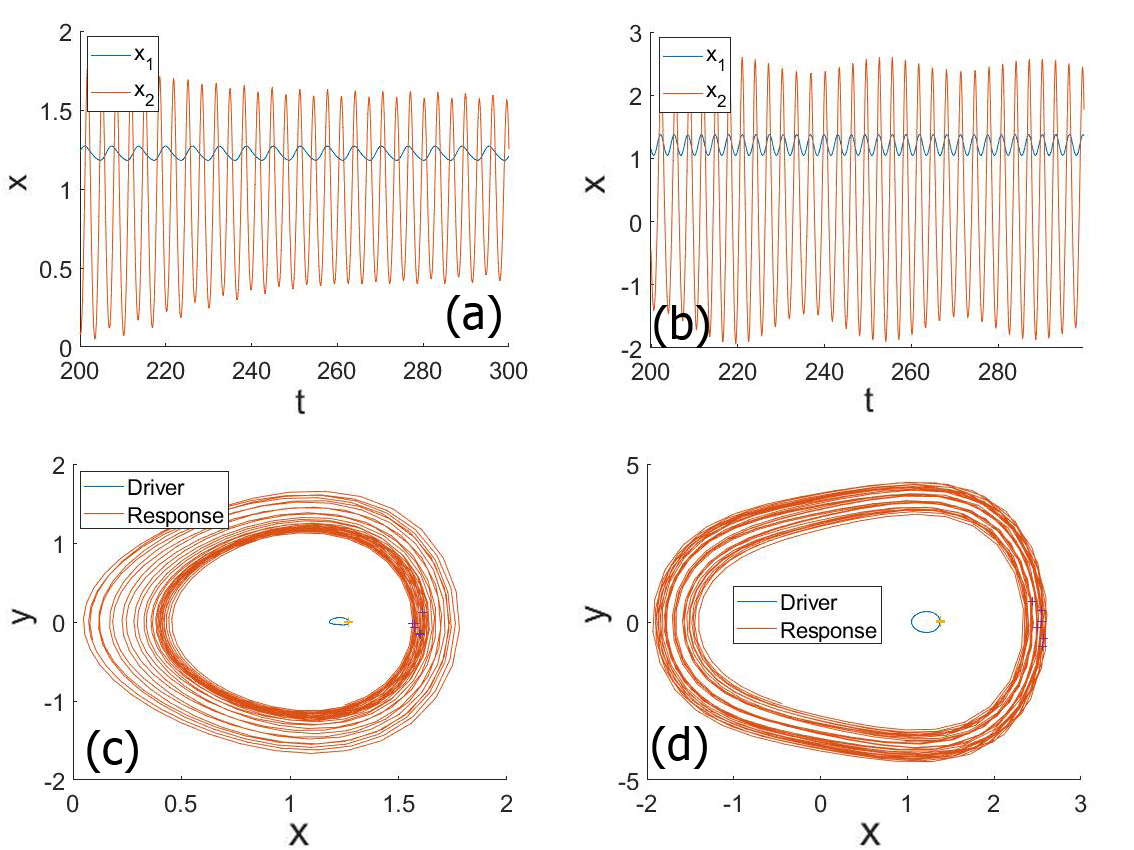}
   \caption{Panels (a) and (b) show the $x-$series and panels (c) and (d) the trajectory in the phase space of the driver and the response system in which we have fixed the forcing amplitude $F=0.1$ for $\tau=1$ and changed the frequency. We have fixed $\omega_1=1$ in panels (a) and (c) and $\omega_1=2$ in panels (b) and (d).}
\label{fig:4}
\end{figure}

\begin{figure}[htbp]
  \centering
   \includegraphics[width=15.0cm,clip=true]{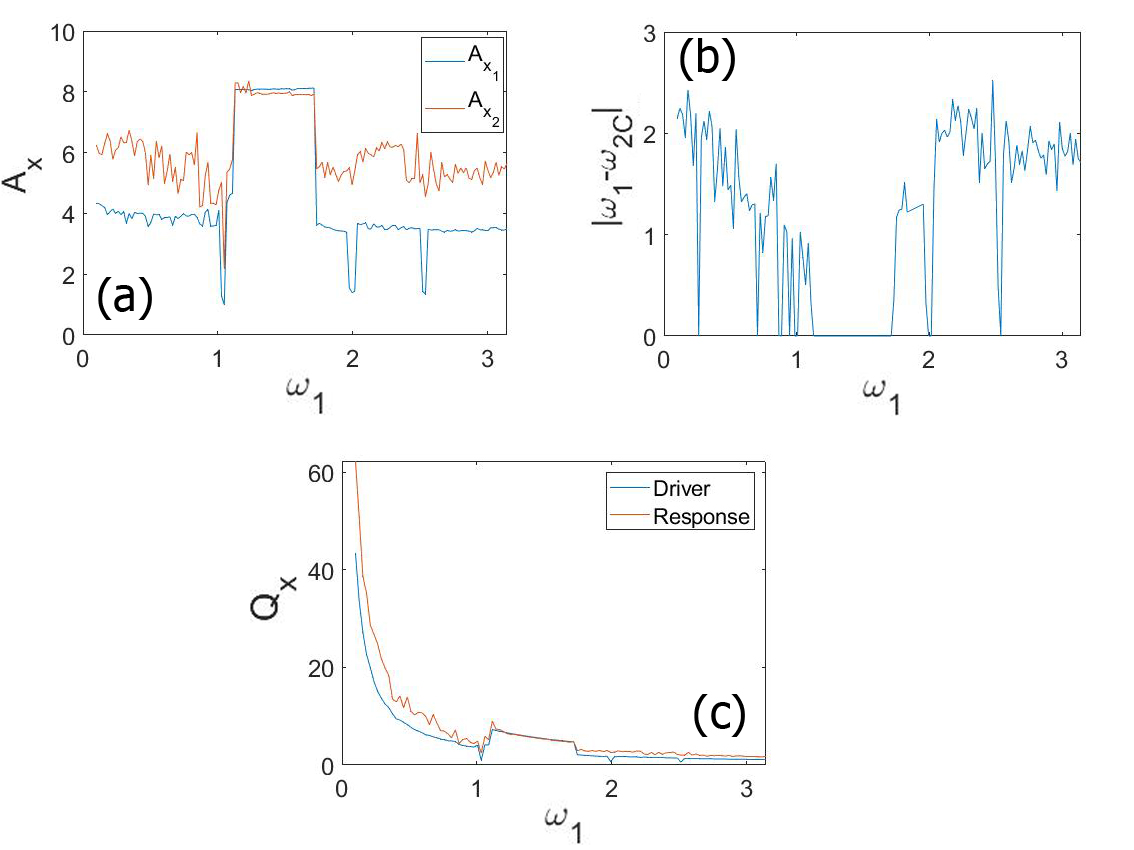}
   \caption{The figure shows in panel (a) the comparison of the oscillations amplitude of the two systems, and in panel (b) the difference of the frequencies of the two oscillators. The frequency of the driver is the forcing frequency while the response system is calculated with the FFT that we have called $\omega_{2C}$. The last panel (c) shows the $Q-$factor for the two systems.}
\label{fig:5}
\end{figure}

\begin{figure}[htbp]
  \centering
   \includegraphics[width=15.0cm,clip=true]{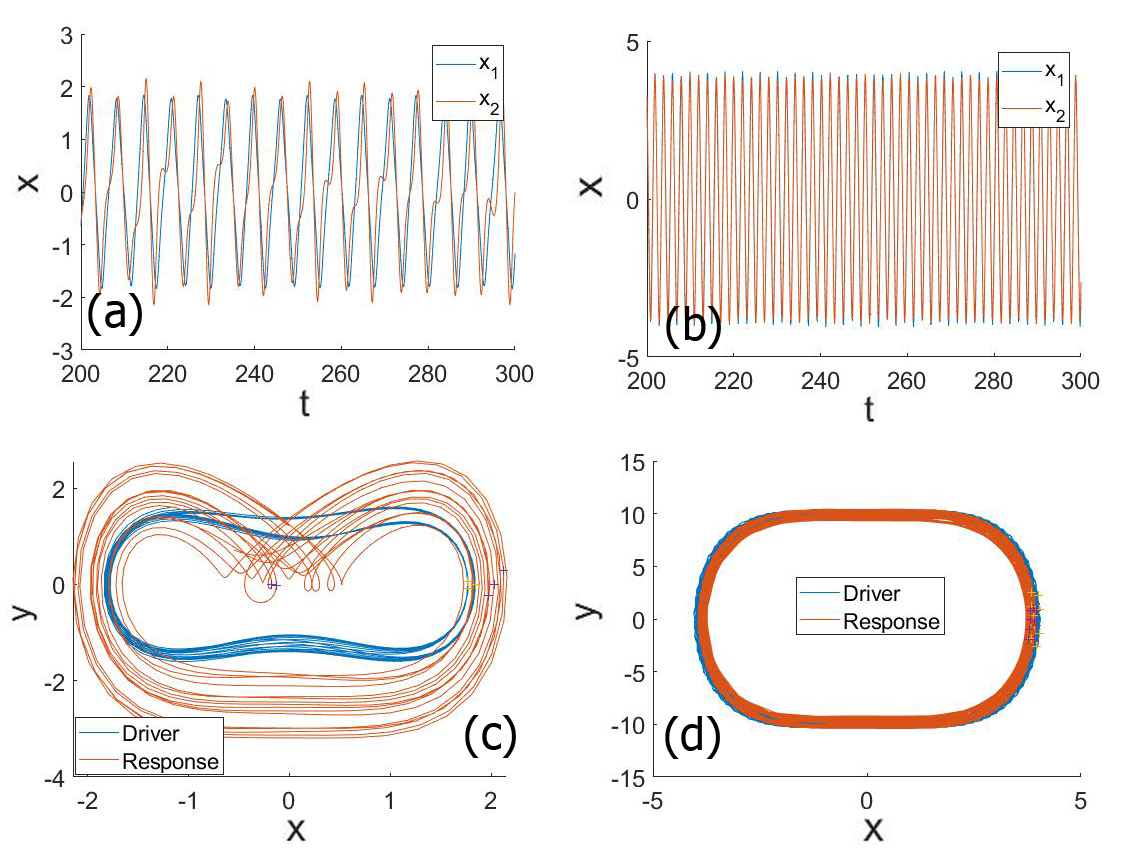}
   \caption{The panels (a) and (b) show the $x-$series and the panels (c) and (d) the trajectory in the phase space of the driver and the response system, where we have fixed the forcing amplitude $F=1$, for $\tau=2$ and varied the frequency. We have fixed $\omega_1=1$ in panels (a) and (c) and $\omega_1=1.5$ in panels (b) and (d).} 
\label{fig:6}
\end{figure}

\begin{figure}[htbp]
  \centering
   \includegraphics[width=16.0cm,clip=true]{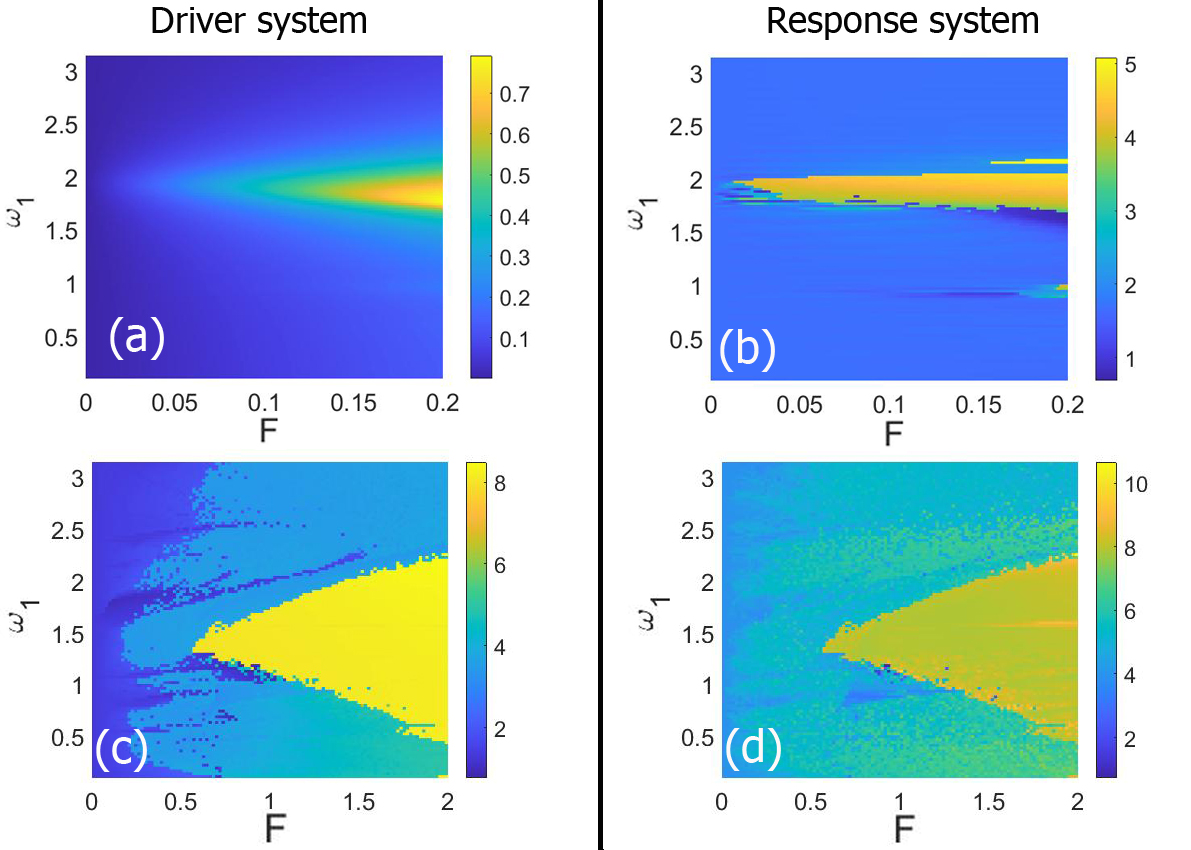}
   \caption{The left panels (a) and (c) show the oscillations amplitude of the driver system, while the right panels (b) and (d) show the oscillations amplitude of the response system. In the first row $\tau=1$ and in the second $\tau=2$. One can observe that the high amplitude oscillations in the driver system are transmitted to the response system. However, in each instance, the oscillation amplitude of the response system is increased through the interaction with its own oscillations. We have fixed $C=1.66$ for all panels.}
\label{fig:7}
\end{figure}

Here, we show that a resonance occurring in the driver system can be transmitted to the response system, through the coupling mechanism. The phenomenon is shown in Fig.~\ref{fig:3} for $\tau=1$. In Fig.~\ref{fig:3}(a), we show the amplitudes of the oscillations of the driver system in function of $\omega_1$ and in Fig.~\ref{fig:3}(b) of the response system. It is possible to check in Fig.~\ref{fig:3}(c) that the driver system resonance has been reproduced in the response system for the same frequency range of $\omega_1$. Important to stress out is that the resonance transmitted from the driver to the response system is magnified in the oscillations of this last one. We will demonstrate that this phenomenon is due to the interaction between the resonance induced by the time-delayed Duffing oscillator for the coupling constant $C=1.66$ and the transmitted driver resonance itself. We can call this phenomenon a {\it superposed resonance}, because implies the superposition of the resonance produced by the external forcing with the previous resonance produced by the interplay of the delay and the coupling. Then, Fig.~\ref{fig:3}(d) shows the difference between the oscillation frequency of the driver and the response system. The frequency of the driver is the forcing frequency and the response system frequency has been calculated with the Fast Fourier Transform and we call it $\omega_{2C}$. We can see that for the values of $\omega_1$ giving birth to the transmitted resonance the difference between the driver and the response system frequency is zero. Finally, Fig.~\ref{fig:3}(e) shows the $Q-$factor evaluated along the variation of the $\omega_1$ parameter of both systems. The $Q-$factor provides an idea of how much the signal is amplified by a certain parameter, in this case case $\omega_1$. We calculate it by computing the sine and cosine components:
\begin{equation}\label{eq:Q}
    B_s=\frac{2}{nT}\int^{nT}_0{x(t)\sin{\omega t}dt}\quad\text{and}\quad B_c=\frac{2}{nT}\int^{nT}_0{x(t)\cos{\omega t}dt},
\end{equation}
where $T=2\pi/\omega$ and $n$ is an integer. Then, we can find the dependence on $\omega_1$ with
\begin{equation}\label{eq:4}
    Q=\frac{\sqrt{B^2_s+B^2_c}}{\omega_1}.
\end{equation}
It shows that the peak of the previous plots is due to a resonance phenomenon generated by the frequency $\omega_1$ in the driver system.
Now, to better understand the phenomenon we plot in Fig.~\ref{fig:4} the $x-$series and the trajectories in the phase space of the systems for two values of $\omega_1$ before $\omega_1=1$ and on the peak $\omega_1=2$. We can see that in the first case the response system trajectory and the $x-$series are confined in one well while in the second case they include both wells. On the other hand, the amplitude of the driver system oscillations are always confined in one well, although in the second case the amplitudes are larger. 
To keep on our analysis, we repeat the same study for $\tau=2$. Therefore,  we show the oscillations amplitude of both systems in Fig.~\ref{fig:5}(a), the difference between the oscillations frequencies in Fig.~\ref{fig:5}(b) and the $Q-$factor in Fig.~\ref{fig:5}(c). The only difference with the $\tau=1$ case is that the oscillations amplitude of the two systems are equivalent. In fact, the $x-$series and trajectories analysis, Fig.~\ref{fig:6}(b) and Fig.~\ref{fig:6}(d), show that the oscillations of both systems encompass the two wells with the same oscillations amplitude.  Here, the  superposed resonance phenomenon is overcast by the already large amplitude of the driver system due to the transmitted resonance. 

Then, in Fig.~\ref{fig:7}, we depict the oscillations amplitude of the driver system and the response system  in the parameter set $\omega_1-F$. Specifically, in Fig.~\ref{fig:7}(a) and Fig.~\ref{fig:7}(c) we show the oscillations amplitude of the driver system.  Then, Fig.~\ref{fig:7}(b) and Fig.~\ref{fig:7}(d) show the amplitude of oscillations in the response system.  It is possible to see that the high oscillations amplitude of the driver system is reproduced in the response system. The distinction lies in the fact that the amplitude of oscillations in the response system surpasses that of the driver system, especially in the case of $\tau=1$, where it reaches an order of magnitude higher.  The transmission phenomenon is shown in Fig.~\ref{fig:7}(b), where the peak of the high oscillations in the driver system around $\omega_1\simeq1.811$ is mirrored by the oscillations in the response system.  The other two figures, Fig.~\ref{fig:7}(c) and Fig.~\ref{fig:7}(d), show the gradient plot of the oscillations amplitude for $\tau=2$.  The response system  replicates the  high amplitude oscillations of the driver system. The calculation of the frequency of the response system oscillations, via the Fast Fourier Transform, that we call $\omega_{2C}$ is worth to analyze. A previous discussion has already been carried out about Fig.~\ref{fig:3} and Fig.~\ref{fig:5}, where we can see on the right column $|\omega_1-\omega_{2C}|$. In each line of Fig.~\ref{fig:7}, we have the frequencies and the frequency difference of the two cases $\tau=1$ and $\tau=2$. The areas where the resonance is transmitted can be identified by observing that $|\omega_1-\omega_{2C}|\rightarrow 0$.
Lastly, we have analyzed the case when a periodic forcing affects also the response system yielding:
\begin{align}\label{eq:5}
Driver&\rightarrow \quad   \frac{d^2x_1}{dt^2}+\mu\frac{dx_1}{dt}+\gamma x_1(t-\tau)+\alpha x_1(1-x_1^2)=F\cos{\omega_1 t},\\
Response&\rightarrow \quad   \frac{d^2x_2}{dt^2}+\mu\frac{dx_2}{dt}+\alpha x_2(1-x_2^2)=C(x_1-x_2)+f\cos{\omega_2 t}. 
\end{align}
We have seen that, once the forcing amplitude $F$ of the driver system is fixed for values for which the resonance is triggered in the driver system, for both $\tau=1$ and $\tau=2$, the phenomenon is robust for values of $f<1$. In both cases, we set two values of the forcing frequency of the response system $\omega_2=0.5$ and $\omega_2=1$ and the results are qualitatively similar. 

\begin{figure}[htbp]
  \centering
   \includegraphics[width=14.0cm,clip=true]{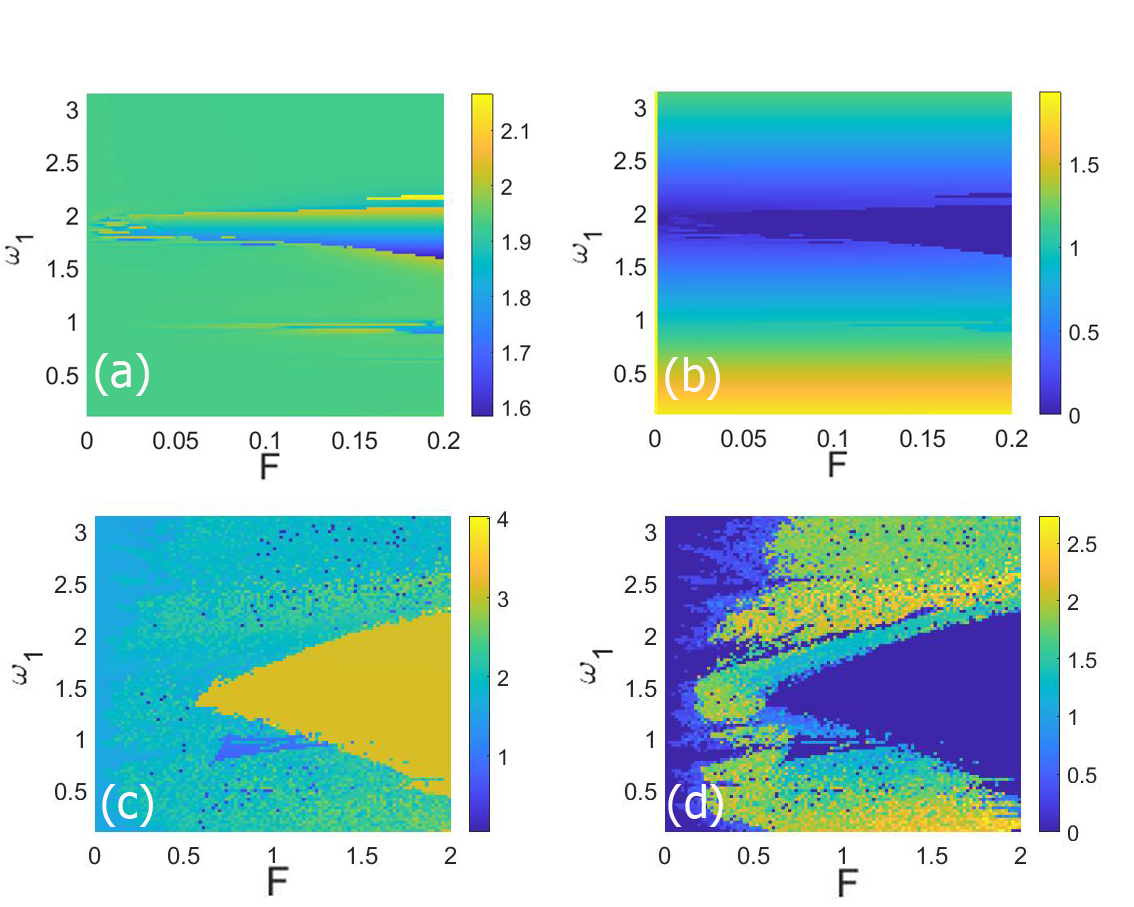}
   \caption{The left panels (a) and (c) show the response system frequency $\omega_{2C}$, calculated with the FFT. The right panels (b) and (d) show the mean difference between the frequency of the driver and the response system. On the first row $\tau=1$ and on the second $\tau=2$.  We can see that in the region in which the resonance has been transmitted, the mean frequency difference tends to zero. }
\label{fig:8}
\end{figure}

\begin{figure}[!htbp]
  \centering
   \includegraphics[width=14.0cm,clip=true]{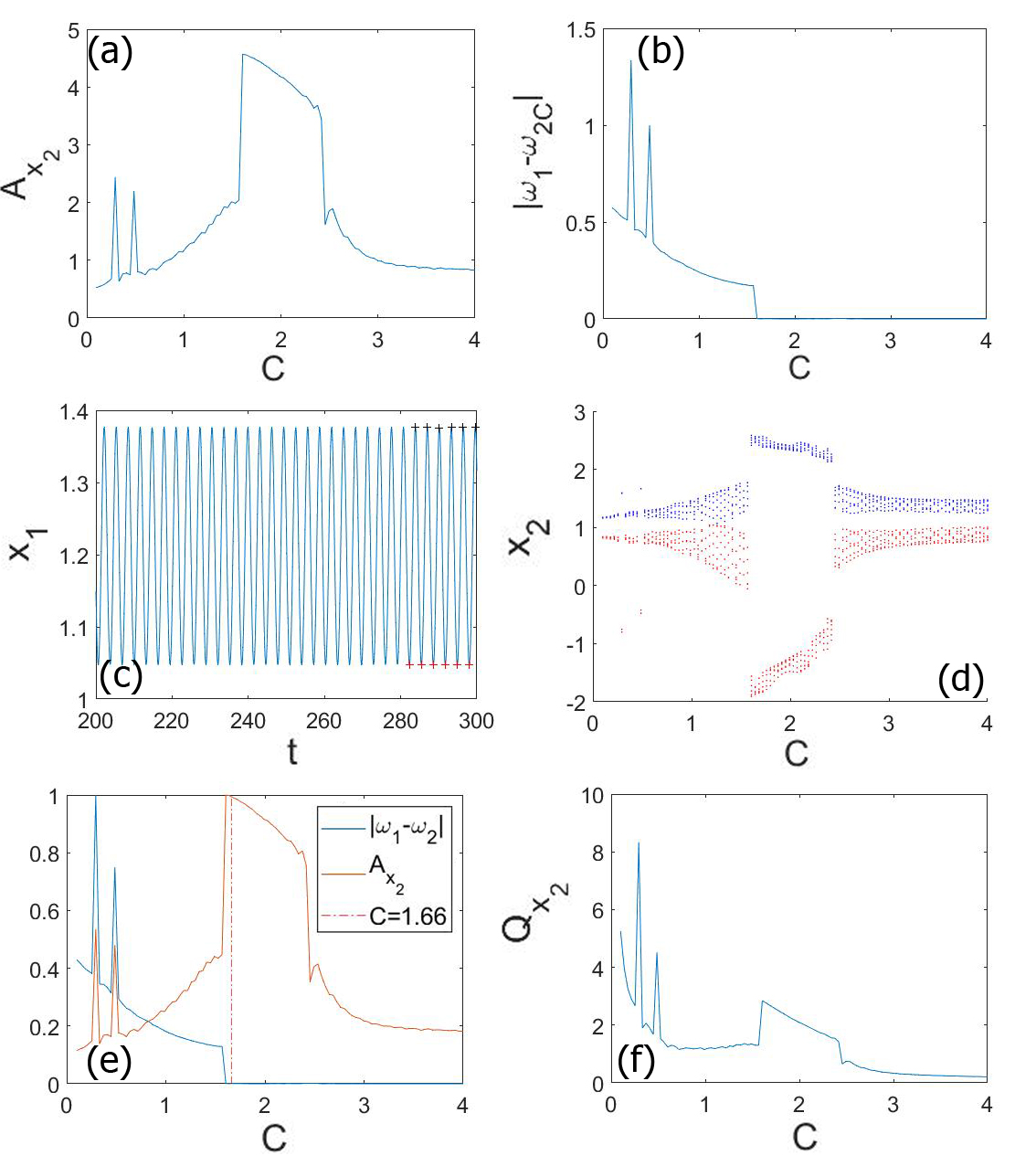}
   \caption{In panel (a) we show the oscillations amplitude of the response system as a function of the coupling constant $C$. In panel (b) the difference of the frequencies of the driver $\omega_1$ and the response system $\omega_{2C}$ as a function of the coupling constant $C$. In panel (c) the $x-$series of the driver system with the points of the Poincar\'e map of the maximum and minimum. In panel (d) the maximum and the minimum of the response system asymptotic oscillations. In panel (e) the comparison of the normalized oscillations amplitude of the response system and the frequencies difference. In panel (f) the $Q$ factor of the oscillations amplitude of the response system in function of the coupling constant $C$. We have fixed the parameters $F=0.1, \tau=1, \omega_1=2$.}
   \label{fig:9}
\end{figure}

\begin{figure}[!htbp]
  \centering
   \includegraphics[width=14.0cm,clip=true]{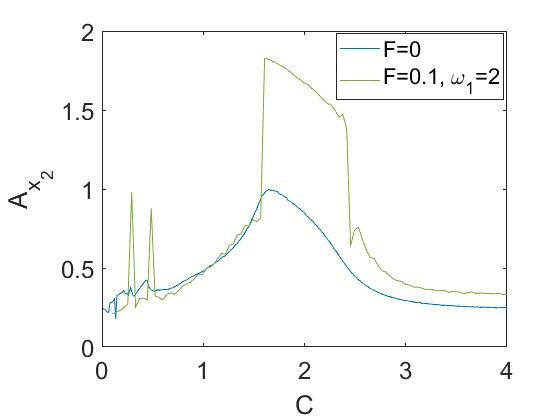}
   \caption{The comparison of the oscillations amplitude curves for $F=0$ and $F=0.1$ and $\omega_1=2$ are shown in the figure. The curve for $F=0.1$ has been reduced in height in order to better appreciate the comparison. Before $C=1.5$ and beyond $C=2.45$ the trend of the two curves are the same. Between those two values of $C$, the transmitted resonance starts and it overlaps with the previous resonance phenomenon due to the interaction between the coupling term and the dynamics of the response system.  }
   \label{fig:10}
\end{figure}

\begin{figure}[!htbp]
  \centering
   \includegraphics[width=14.0cm,clip=true]{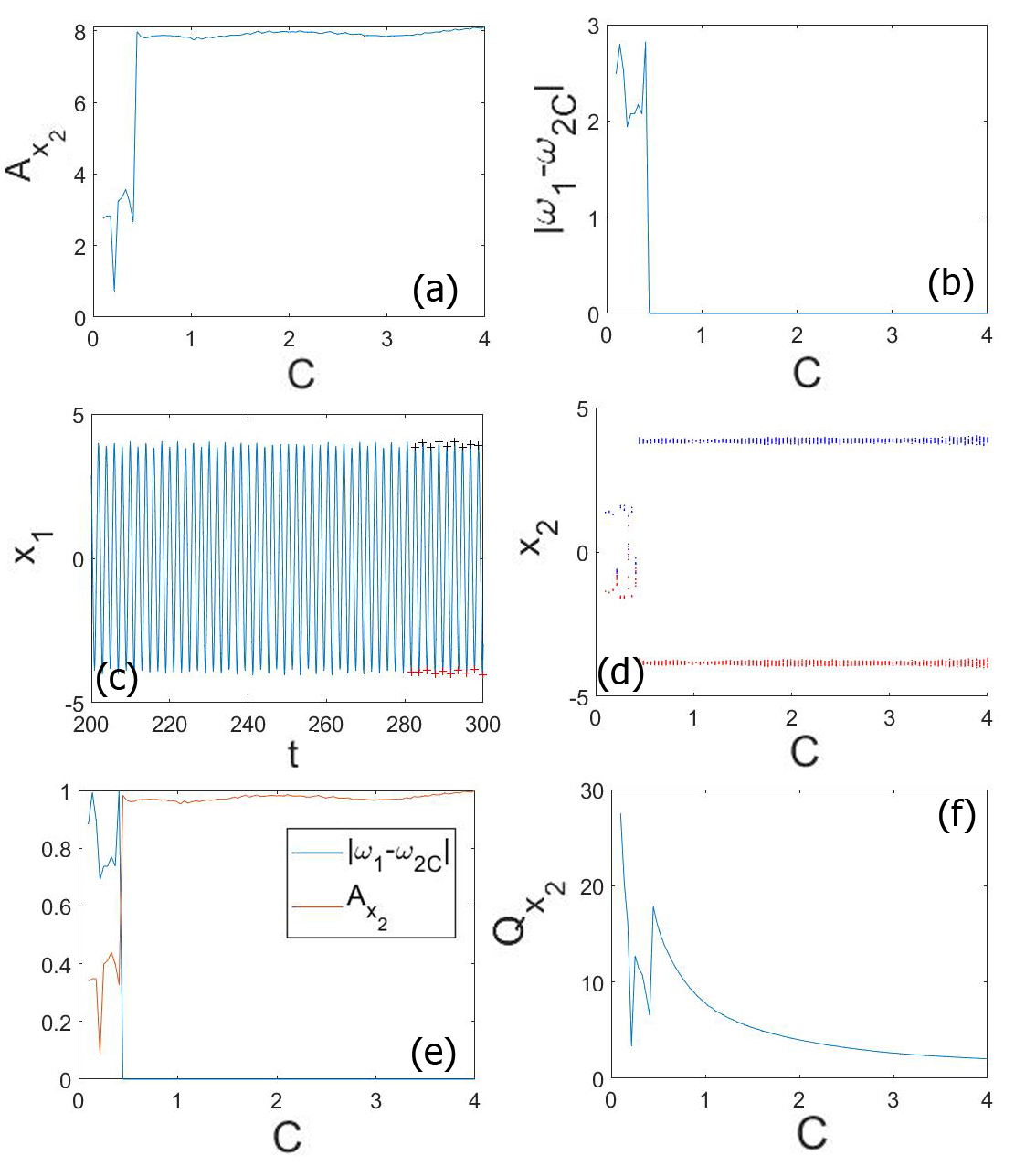}
   \caption{In panel (a) we show the oscillations amplitude of the response system in function of the coupling constant $C$. In panel (b) the difference of the frequencies (b) of the driver $\omega_1$ and the response system $\omega_{2C}$  in function of the coupling constant $C$. In panel (c) the $x-$series of the driver system with the points of the Poincar\'e map of the maximum and minimum. In panel (d) the maximum and the minimum of the response system asymptotic oscillations. in panel (e) the comparison of the normalized oscillations amplitude of the response system and the frequencies difference. In panel (f) the $Q$ factor of the oscillations amplitude of the response system in function of the coupling constant $C$. We have fixed the parameters $F=1, \tau=2, \omega_1=1.5$. }
   \label{fig:11}
\end{figure}

\begin{figure}[!htbp]
  \centering
   \includegraphics[width=15.0cm,clip=true]{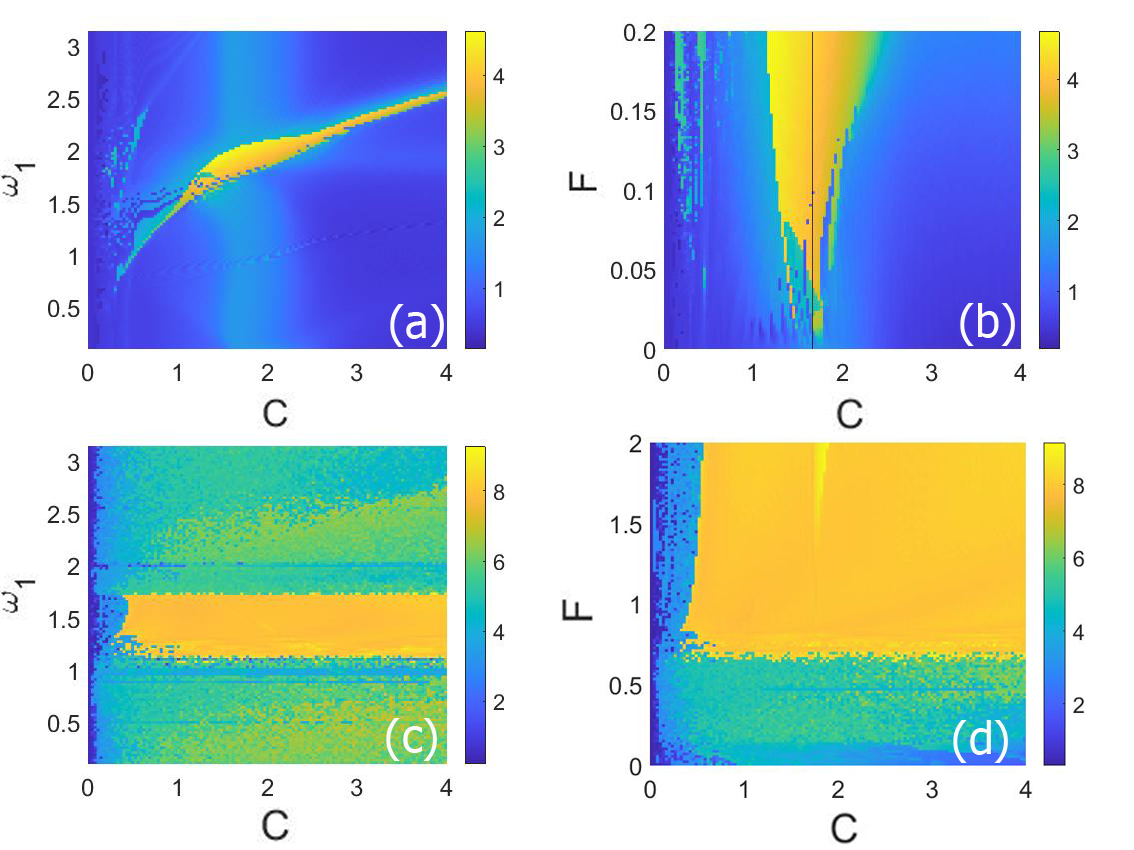}
   \caption{The figure shows in panel (a) the oscillations amplitude of the response system as a function of $\omega_1$ and $C$ for $\tau=1, F=0.1$ and in panel (c) for $\tau=2, F=1$. Then, in panel (b)  the oscillations amplitude as a function of $F$ and $C$ for $\tau=1, \omega_1=1.881$ and in panel (d) for $\tau=2, \omega_1=1.5$.}
   \label{fig:12}
\end{figure}

\section{Effect of the coupling constant on the transmitted resonance}\label{Sec:IV}

The $C$ choice adopted until now has been already discussed. Nevertheless, we decided to analyze the role of the coupling constant in the transmitted resonance. So, we depict Fig.~\ref{fig:9} and Fig.~\ref{fig:11}. Here, we show the oscillations amplitude of the response system, the frequencies difference $|\omega_1-\omega_{2C}|$, the $x-$series of the driver system, the maximum-minimum diagram of the response system, the comparison of the oscillations amplitude and the frequencies difference and the $Q-$factor in function of $C$ for $\tau=1$ and $\tau=2$, respectively. In the first figure, Fig.~\ref{fig:9} ($\tau=1$), we can see that there are optimal values of the coupling constant $1.5<C<2.5$ for which the oscillations amplitude is larger. Moreover, the $x-$series of the driver system Fig.~\ref{fig:9}(c) and the maximum-minimum diagram Fig.~\ref{fig:9}(d) of the response system indicate that for those optimal $C$ values there is a huge  difference in the behavior of the two system. If we compare Fig.~\ref{fig:9}(b) and Fig.~\ref{fig:9}(e), we can observe that when $C$ reaches the first optimal value, the difference in frequency drops to $0$. Finally, the $Q-$factor shown in Fig.~\ref{fig:9}(f), calculated along the variation of the parameter $C$, shows that those optimal $C$ values trigger a resonance over the transmitted resonance, what we call a  superposed resonance. This  superposed resonance gives rise to a much larger oscillations amplitude. In fact, in Fig.~\ref{fig:10} we compare the oscillations amplitude shown in Fig.~\ref{fig:9}(a) with the oscillations amplitude of the response system for $F=0,\tau=1$. Subsequently, it becomes evident that the curves exhibit a consistent pattern for values of \(C\) below 1.5 and above 2.5. However, within this range, a substantial disparity emerges, attributed to the initiation of the transmitted resonance and its confluence with a preceding resonance phenomenon.  This explains the high oscillations amplitude compared with the driver system in Fig.~\ref{fig:7}.
On the other hand, Fig.~\ref{fig:11}(a) ($\tau=2$), shows a threshold value for which the transmitted resonance starts $C\approx0.5$ and then the oscillations amplitude stabilizes for higher $C$ values. If we compare Fig.~\ref{fig:11}(b) and Fig.~\ref{fig:11}(e), we can observe that when $C$ reaches the first optimal value, the difference in frequency drops to $0$. The difference is that the oscillations amplitude of the response system beyond the $C$ threshold value are comparable with the oscillations amplitude of the driver system, see Fig.~\ref{fig:11}(c) and Fig.~\ref{fig:11}(d). Indeed, the $Q$-factor does not indicate that the coupling term initiates a resonance beyond the transmitted resonance, as depicted in Fig.~\ref{fig:11}(f). The only discernible peak aligns precisely with the onset of the transmitted resonance. Hence, the previously noted observation is affirmed here: the already broad oscillations resulting from the transmitted resonance mask the resonance phenomenon arising from the interplay between the dynamics of the response system and the coupling term. This elucidates why, for this particular value of $\tau=2$, the amplitude of oscillations in both the driver and the response system resembles that in Fig.~\ref{fig:7}.
Finally, the study of the oscillations amplitude in the parameter sets $\omega_1-C$ and $F-C$ is carried out in Fig.~\ref{fig:12}. The Fig.~\ref{fig:12}(a) and Fig.~\ref{fig:12}(c) shows very different behaviors of this resonance in the two $\tau$ regions. In fact, for $\tau=1$, Fig.~\ref{fig:12}(a), the frequency $\omega_1$ values for which the transmitted resonance is triggered vary in function of $C$. On the other hand, for $\tau=2$ the frequency values $\omega_1$ triggering the transmitted resonance are always the same for all the $C$ values, as can be seen in Fig.~\ref{fig:12}(c).

Then, Fig.~\ref{fig:12}(b) shows that there is a particular range of values of $C$ giving birth to the transmitted resonance that varies as a function of $F$. On the other hand, Fig.~\ref{fig:12}(d) shows that there is a threshold value, $F=0.64$, that generates the resonance in the driver, and a minimum value, $C=0.31$, for which the transmission  of the resonance starts.

\section{Conclusions}\label{Sec:concl}

The study of two coupled systems has been carried out. The driver is a time-delayed Duffing oscillator, while the response system is a Duffing oscillator with no delay. The first one is the forcing of the response system. Then, a periodic forcing affects the driver system. The goal of this work is  to study the transmission of a driver system resonance generated by its own periodic forcing from the driver system to the response system. Here, we find the conditions that make this possible. Moreover, this transmitted resonance interacts with a previous resonance, triggered by the driver system as the sole forcing of the response system, through the coupling constant. The effect is that the interplay between the two resonances magnifies the amplitude of the oscillations in the driver system transmitted to the response system.  We call this new phenomenon  superposed resonance. The last result is that the high oscillations amplitudes of the  superposed resonance are transmitted to the response system with their frequencies, so that the response system oscillates with the same frequency as the driver system.

The phenomenon of the transmitted resonance has several potential practical and physical applications across various fields. \textit{Signal Transmission and Amplification}: Transmitted resonance can be utilized in communication systems to enhance the transmission of signals between coupled systems. By exploiting the superposed resonance effect, it becomes possible to amplify and transmit signals with higher amplitudes and frequencies. \textit{Synchronization of Oscillators:} In systems where synchronization of oscillators is desirable, such as in electronic circuits or biological systems, the transmission of resonance between coupled systems can facilitate the synchronization at specific frequencies. This synchronization can improve the efficiency and performance of such systems. \textit{Vibration Control and Damping:} Understanding the transmitted resonance can aid in the design of vibration control systems for structures prone to oscillations or vibrations. By selectively transmitting resonant frequencies from one system to another, it becomes possible to control and dampen unwanted vibrations, thereby improving structural stability and safety. \textit{Energy Harvesting:} In energy harvesting systems, the ability to transmit resonance between coupled oscillators can enhance the efficiency of energy conversion. By exploiting resonant frequencies, it becomes possible to harvest energy more effectively from ambient vibrations or mechanical sources. \textit{Sensor Networks: }Transmitted resonance can also find applications in sensor networks where multiple sensors need to communicate and synchronize their measurements. By leveraging the superposed resonance phenomenon, sensor networks can achieve synchronized operation and improve the accuracy and reliability of measurements. Overall, the practical and physical applications of transmitted resonance are diverse and span various fields, including communication systems, synchronization, vibration control, energy harvesting, and sensor networks. Further research and development in this area can lead to innovative solutions and advancements in these domains.

\section{Acknowledgment}
This work has been supported by the Spanish State Research Agency (AEI) and the European Regional Development Fund (ERDF, EU) under Project No.~PID2019-105554GB-I00.


\end{document}